\documentclass[12pt]{article}

\newcommand{\Tr}{\rm{Tr}}

\usepackage{latexsym,amsfonts,amsmath,amsthm,amssymb}

\usepackage[T1]{fontenc}
\usepackage[latin1]{inputenc}
\title{Asymptotic expansion of Gaussian integrals of analytic functionals on infinite-dimensional spaces
and quantum averages}
\author{Andrei Khrennikov\\
International Center for Mathematical
Modeling \\ in Physics and Cognitive Sciences,\\
University of V\"axj\"o, S-35195, Sweden}

\begin{document}
\maketitle

\abstract{We study asymptotic expansions of Gaussian integrals of
analytic functionals on infinite-dimensional spaces (Hilbert and
nuclear Frechet). We obtain an asymptotic equality coupling the
Gaussian integral  and the trace of the composition of scaling of
the covariation operator of a Gaussian measure and the second
(Frechet) derivative of a functional. In this way we couple
classical average (given by an infinite-dimensional Gaussian
integral) and quantum average (given by the von Neumann trace
formula). We can interpret this mathematical construction as a
procedure of ``dequantization'' of quantum mechanics. We represent
quantum mechanics as an asymptotic projection of classical
statistical mechanics with infinite-dimensional phase-space. This
space can be represented as the space of classical fields, so
quantum mechanics is represented as a projection of ``Prequantum
Classical Statistical Field Theory''.}

\section{Introduction}

The problem of reduction of  quantum mechanics to classical
statistical mechanics has been discussed from the first days of
quantum mechanics, see, e.g., [1]--[45]. Now days this problem is
known as the problem of hidden variables or completeness of quantum
mechanics, see, e.g., [33] --[37] for recent debates. There is  a
rather common opinion that quantum mechanics is complete and that it
is impossible to introduce ``hidden variables'' providing more
detailed description than quantum mechanics. But in  [46] it was
demonstrated that in the opposition to this opinion it is possible
to represent quantum mechanics as a projection of classical
statistical mechanics on {\it infinite-dimensional space.} In this
paper we present this approach (which was called in [46] {\it
Prequantum Classical Statistical Field Theory} -- PCSFT) on the
mathematical level of rigorousness; in particular, some functional
spaces introduced in [46] should be modified to obtain the correct
results; moreover, in the basic asymptotic equality coupling
classical and quantum averages we obtain an estimate of the rest
term, $o(\alpha).$

 In the present paper we also find connection of
PCSFT with background Gaussian random field on Hilbert space.
Finally, we extend the approach of [46] by considering unbounded
operators, see section 6. There are also differences in
interpretations of a small parameter of our asymptotic procedure of
{\it dequantization.} In [46] this parameter was identified with the
Planck constant $h$ (I was very much stimulated by discussions with
people working in SED and stochastic quantum mechsnics).  In this
paper we introduce a new parameter $\alpha$ giving the dispersion of
prequantum fluctuations, see [47] for more details on physical
interpretation. To simplify considerations, in  this paper we
consider quantum formalism over the field of real numbers, see [47]
for complex theory. To exclude possible misunderstanding, we
emphasize from the very beginning that our paper {\it  is not about
deformation quantization for systems with the infinite number of
degrees of freedom,} see, e.g., [48], [49], but about dequantization
of conventional quantum mechanics for systems with  a finite number
of degrees of freedom by means of analysis on infinite-dimensional
space.

Our model is classical statistical mechanics on the phase space
$\Omega= H\times H,$ where $H$ is the real  Hilbert space. Points of
this phase-space can be considered as {\it classical fields} (if we
take the Hilbert space $H=L_2({\bf R}^3)).$ Our approach can be
called {\it Prequantum Classical Statistical Field Theory} - PCSFT.

Our approach is an asymptotic approach. We introduce a small
parameter $\alpha$ -- dispersion of ``vacuum fluctuations''. In fact
we consider a one parameter family of classical statistical models
$M^\alpha.$ QM is obtained as the limit of classical statistical
models when $\alpha \to 0:$
\begin{equation} \label{CP}
 \lim_{\alpha\to 0}M^\alpha=N_{\rm{quant}},
\end{equation}
where $N_{\rm{quant}}$ is the Dirac-von Neumann quantum model [2],
[4]. As was already remarked, our approach should not be mixed with
so called {\it deformation quantization.} In the formalism of
deformation quantization classical mechanics on the phase-space
$\Omega_{2n} = {\bf R}^{2n}$ is obtained as the $\lim_{h\to 0}$ of
quantum mechanics (the correspondence principle). In the deformation
quantization the quantum model is considered as depending on a small
parameter $h:N_{\rm{quant}}\equiv N_{\rm{quant}}^h,$ and formally
\begin{equation}
\label{CP1}
 \lim_{h \to 0} N_{\rm{quant}}^h = M_{\rm{conv. class.}}
\end{equation}
where  $M_{\rm{conv. class.}}$ is the conventional classical model
with the phase-space $\Omega_{2n}.$

In our approach the classical $\to $ quantum correspondence $T$ is
based on the Taylor expansion of classical physical variables --
functions $f: \Omega \to {\bf R}.$ This is a very simple map:
function is mapped into its second derivative (which is always a
symmetric operator).\footnote{By the terminology which is used in
functional analysis $f$ is called functional -- a map from a
functional space into real numbers. If we represent $\Omega$ as the
space of classical fields, $\psi:{\bf R}^3 \to {\bf R},$ then
$f(\psi)$ is a functional of classical field.}

The space of classical statistical states consists of Gaussian
measures on $\Omega$ having zero mean value and dispersion $\approx
\alpha.$ Thus a statistical state $\rho$ (even a so called pure
state $\psi\in \Omega, \Vert \psi \Vert=1)$ can be interpreted as a
Gaussian ensemble of classical fields which are very narrow
concentrated near the vacuum field $\psi_{\rm{vacuum}}(x)=0$ for all
$x \in {\bf R}^3.$ Such a $\rho$ has the very small standard
quadratic deviation from the field of vacuum $\psi_{\rm{vacuum}}:$
\begin{equation}
\label{SD} \int_{L_2({\bf R}^3)
 \times L_2({\bf R}^3)} \int_{{\bf
R}^3} [p^2(x) + q^2(x)] dx d\rho (q,p) = \alpha, \; \alpha\to 0,
\end{equation}
where a classical (prequantum) field $\psi(x)$ is a vector field
with two components $ \psi(x)=(q(x), p(x)).$ The field has the
dimension of energy  per volume (as in the case of electromagnetic
field in the Gaussian system of units).\footnote{So we really
interpret $\psi$ as a classical field and not as a square root of
probability.}  Then a statistical state $\rho$ is an ensemble of
fluctuations of vacuum which are small in the energy domain.

The choice of the space of statistical states plays the crucial role
in our approach. QM is the image of a very special class of
classical statistical states. Therefore we discuss this problem in
more detail. Let us use the language of probability theory. Here a
statistical state is represented by a Gaussian random variable $
\lambda \to \psi_\lambda,$ where $\lambda$ is a random parameter. We
have:
\begin{equation}
\label{SD1} E \psi_\lambda=0, \sigma^2(\psi)=  E \vert \psi_\lambda
- \psi_{\rm{vacuum}} \vert^2  =\alpha.
\end{equation}
We pay attention to the evident fact that small dispersion does not
imply that the random variable $\psi_\lambda$ is small at any point
$\lambda \in \Lambda,$ where $\Lambda$ is the space of random
parameters. Here smallness is considered with respect to the
$L_2$-norm. The internal energy of the field $\psi_\lambda,$
$$
{\cal E} (\psi_\lambda)\equiv \int_{{\bf R}^3} \vert \psi_\lambda(x)
\vert^2 d x = \int_{{\bf R}^3} [p^2_\lambda (x) + q^2_\lambda(x)]
dx,
$$
can be arbitrary large (with nonzero probability). But the
probability that ${\cal E} (\psi_\lambda)$ is sufficiently large is
very small. The easiest way to estimate this probability is to use
the (well known in elementary probability theory) Chebyshov
inequality:
\begin{equation}
\label{SD2} P( \lambda : {\cal E} (\psi_\lambda) > C) \leq E {\cal
E}(\psi_\lambda)/C= \alpha/C \to 0, \alpha\to 0,
\end{equation}
for any constant $C>0.$

It is especially interesting  that in our approach ``pure quantum
states'' are not pure at all! These are also statistical mixtures of
small Gaussian fluctuations of the ``background field''.

At the moment we are not able to estimate the magnitude $\alpha$ of
Gaussian vacuum fluctuations. In the first version of our work [46]
we assumed, as it is common in SED [50], [51] as well as in
stochastic QM [18], that $\alpha$ has the magnitude of  the Planck
constant $h.$ However, we could not justify this fundamental
assumption on the magnitude of vacuum fluctuations in our approach.
It may be that vacuum fluctuations described by PCSFT are
essentially smaller than fluctuations considered in  SED and 
stochastic QM. One might even speculate on a connection with {\it
cosmology and string theory.} However, in the present paper we
consider the magnitude of vacuum fluctuations just as  a small
mathematical parameter of the model: $\alpha \to 0.$

After publication of paper [1],  I was informed about the paper of
Alexander Bach [52] (see also earlier publications [53], [54]) who
also used the Hilbert phase space to construct a classical
probabilistic representation of quantum mechanics. In both
approaches there was used the representation of the von Neumann
trace formula for quantum averages through integration on the
Hilbert phase space. In this sense my approach is a natural
development of Bach's approach [52], [53], [54]. However, a new
contribution was really nontrivial. Therefore Alexander Bach and I
finally came to completely different conclusions on the possibility
to reduce quantum mechanics to classical statistical mechanics. We
recall the main conclusion of A. Bach [52], p.128 : ``Although we
give a representation of quantum mechanics in terms of classical
probability theory, {\it the concepts of classical probability
theory are not appropriate for quantum theory}.''   My main
conclusion is completely opposite: quantum mechanics can be
represented in a natural way as an approximation of {\it statistical
mechanics of classical fields.}

The main difference between my theory (which was called Prequantum
Classical Statistical Field Theory -- PCSFT) and Bach's theory is
the asymptotic approach to correspondence between the classical and
quantum statistical models. In PCSFT there is a small parameter
$\alpha$ giving the magnitude of fluctuations in Gaussian ensembles
of classical fields.\footnote{We remark that we consider
fluctuations in ensembles of classical fields, but not fluctuations
of a single field on physical space ${\bf R}^3.$} We consider not
only quadratic functions of fields, but arbitrary smooth functions
(as in the classical statistical mechanics). Quantum observables are
obtained through expansion of such functions into the Taylor series.

This viewpoint to quantum mechanics -- as the second order
approximation of classical statistical mechanics on Hilbert phase
space -- gives the possibility to solve a problem that was crucial
for Bach's model (and as we see from his article [52] that problem
was the main reason for rather pessimistic Bach's conclusion which
was mentioned above). This is the problem of correspondence between
functions of classical physical variables and functions of
operators. If $f$ is a classical variable (in our approach an
arbitrary smooth function on the Hilbert phase space and in Bach's
approach a quadratic form) and $T(f)$ is the corresponding quantum
observable (a self-adjoint operator), then, for example,
\begin{equation}
\label{VC} T(f^2) \not= T(f)^2.
\end{equation}
This is not a purely mathematical problem. As was pointed out by
Alexander Bach, this is the root of difference in the definition of
dispersion free states in  the quantum model and in a prequantum
classical statistical model with the Hilbert phase space.

It was totally impossible to solve this problem in Bach's framework.
His prequantum model was an exact one. Therefore the violation of
the equality $T(f^2) = T(f)^2$  was considered as the evidence of
inadequacy of this model to quantum mechanics. Our model, PCSFT, is
not an exact model. This is an asymptotic model or better to say an
a prequantum model which is approximated by the quantum model.

However, {\it the violation of, e.g., the equality $T(f^2) = T(f)^2$
in some   approximation scheme  was never considered in
approximation theory as an evidence of inadequacy of this scheme}.
For example, let us consider the approximation of smooth functions
$f: {\bf R}^n \to {\bf R}$ by their Taylor polynomials of the order
$m.$ This approximation scheme induces the map
\begin{equation}
\label{VC} T: C^\infty \to {\cal P}_m,
\end{equation}
where $C^\infty$ and ${\cal P}_m$ are, respectively, spaces of
smooth functions and polynomials of the degree $m.$ Then it is
evident that (as in our prequantum model), e.g., the equality
$T(f^2) = T(f)^2$ can be violated. But nobody would conclude that
physics described by polynomials of the degree $m$ (e.g., $m=2)$
differs crucially from physics described by smooth functions.

We finish our comparative analysis with Bach's model by emphasizing
that PCSFT  provides the natural interpretation of  hidden
variables: these are classical fields. But in [52] there was still
pointed out that ``... the fact that elements of Hilbert space have
no empirical meaning indicates that the theory still remains open to
interpretations.''

We also consider generalizations of quantum formalism based on
expansions of functionals of classical fields into the Taylor series
up to terms of degree $n,$ see section 7; for $n=2$ we obtain the
conventional quantum mechanics.

\section{Infinite-dimensional analysis}

Gaussian stochastic analysis on infinite-dimensional spaces is a
well established mathematical formalism, see, e.g., Skorohod [55]
for introduction, see [56], [57], [58] for more detail (especially
for applications in sections 5.2 and 6). We also pay attention that
Gaussian analysis on infinite-dimensional spaces was used a lot in
Euclidean quantum field theory, see, e.g., [59], [60].

Let $H$ be a real Hilbert space and let $A: H \to H$ be a continuous
self-adjoint linear operator. The basic mathematical formula which
will be used in this paper is the formula for a Gaussian integral of
a quadratic form $f(\psi)\equiv f_A (\psi)= (A\psi, \psi).$

Let $d\rho(\psi)$ be a $\sigma$-additive Gaussian measure on the
$\sigma$-field $F$ of Borel subsets of $H,$ see [56]--[59]. This
measure is determined by its covariation operator $B: H\to H$ and
mean value $m\equiv m_\rho \in H.$ For example, $B$ and $m$
determines the Fourier transform of $\rho:$ $ \tilde \rho (y)=
\int_H e^{i(y, \psi)} d\rho (\psi)= e^{\frac{1}{2}(By, y) + i(m,
y)}, y \in H.$ In what follows we restrict our considerations to
Gaussian measures with zero mean value $m=0,$ where $ (m,y) = \int_H
(y, \psi) d\rho (\psi)= 0$ for any $y \in H.$ Sometimes there will
be used the symbol $\rho_B$ to denote the Gaussian measure with the
covariation operator $B$ and $m=0.$ We recall that the covariation
operator $B\equiv \rm{cov} \; \rho$ is defined by $(By_1, y_2)=\int
(y_1, \psi) (y_2, \psi) d\rho(\psi), y_1, y_2 \in H,$ and has the
following properties: a). $B \geq 0,$ i.e., ($By, y) \geq 0, y \in
H;$ b). $B$ is a self-adjoint operator, $B \in {\cal L}_{{\rm
s}}(H);$ c). $B$ is a trace-class operator and ${\rm Tr}\; B=\int_H
||\psi||^2 d\rho(\psi).$ This is {\it dispersion} $\sigma^2(\rho)$
of the probability $\rho.$ Thus $\sigma^2(\rho)= {\rm Tr}\; B.$

We pay attention that the list of properties of the covariation
operator of a Gaussian measure differs from the list of properties
of a von Neumann density operator [4] only by one condition:
$\rm{Tr} \; D =1,$ for a density operator $D.$

We can easily find the Gaussian integral of the quadratic form
$f_A(\psi):$
\begin{equation}
\label{QI} \int_H f_A(\psi) d\rho (\psi)={\rm Tr}\; BA
\end{equation}

The differential calculus for maps $f: H\to {\bf R}$ does not differ
so much from the differential calculus in the finite dimensional
case, $f: {\bf R}^n \to {\bf R}.$ Instead of the norm on ${\bf
R}^n,$ one should use the norm on $H.$ We consider so called Frechet
differentiability. Here a function $f$ is differentiable if it can
be represented as $ f(\psi_0 + \Delta \psi)= f(\psi_0) +
f^\prime(\psi_0)(\Delta \psi) + o(\Delta \psi),$ where $\lim_{\Vert
\Delta \psi \Vert\to 0} \frac{\Vert o(\Delta \psi)\Vert }{\Vert
\Delta \psi \Vert} =0.$ Here at each point $\psi$ the derivative
$f^\prime(\psi)$ is a continuous linear functional on $H;$ so it can
be identified with the element $f^\prime(\psi)\in H.$ Then we can
define the second derivative as the derivative of the map $\psi\to
f^\prime(\psi)$ and so on. A map $f$ is differentiable $n$-times
iff: $$ f(\psi_0 + \Delta \psi)= f(\psi_0) + f^\prime(\psi_0)(\Delta
\psi) + \frac{1}{2}f^{\prime \prime}(\psi_0)(\Delta \psi, \Delta
\psi) + ...
$$
$$
+\frac{1}{n!} f^{(n)}(\psi_0)(\Delta \psi, ...,
\Delta \psi)+ o_n(\Delta \psi),$$
where $f^{(n)}(\psi_0)$ is a
symmetric continuous $n$-linear form on $H$ and $ \lim_{\Vert \Delta
\psi \Vert\to 0} \frac{\Vert o_n(\Delta \psi)\Vert }{\Vert \Delta
\psi \Vert^n} =0. $ For us it is important that
$f^{\prime\prime}(\psi_0)$ can be represented by a symmetric
operator $f^{\prime \prime}(\psi_0)(u,v)=(f^{\prime \prime}(\psi_0)
u, v), u, v \in H $ (this fact is well know in the finite
dimensional case: the matrix representing the second derivative of
any two times differentiable function $f: {\bf R}^n \to  {\bf R}$ is
symmetric).
 We remark that in this case $f(\psi)= f(0) +  f^\prime(0)(\psi) + \frac{1}{2}f^{\prime
\prime}(0)(\psi, \psi) + ...+ \frac{1}{n!} f^{(n)}(0)(\psi,
...,\psi) + o_n(\psi).$

For a real Hilbert space $H,$ denote by the symbol $H^{{\bf C}}$ its
complexification: $H^{{\bf C}}= H\oplus i H.$ We recall that a
function $f: H^{{\bf C}} \to {\bf C}$ is analytic if it can be
expanded into the Taylor series:
\begin{equation}
\label{ANN} f(\psi)= f(0) +  f^\prime(0)(\psi) +
\frac{1}{2}f^{\prime \prime}(0)(\psi, \psi) + ...+ \frac{1}{n!}
f^{(n)}(0)(\psi, ...,\psi) +... .
\end{equation}
which converges uniformly on any ball of $H^{{\bf C}}.$

\section{Dequantization}

\subsection{Classical and quantum  statistical models}
We define {\it ``classical statistical models''} in the following
way, see [46] for more detail (and even philosophic considerations):
a) physical states $\omega$ are represented by points of some set
$\Omega$ (state space); b) physical variables are represented by
functions $f: \Omega \to {\bf R}$ belonging to some functional space
$V(\Omega);$ c) statistical states are represented by probability
measures on $\Omega$ belonging to some class $S(\Omega);$ d) the
average of a physical variable (which is represented by a function
$f \in V(\Omega))$ with respect to a statistical state (which is
represented by a probability measure  $\rho \in S(\Omega))$ is given
by
\begin{equation}
\label{AV0} < f >_\rho \equiv \int_\Omega f(\psi) d \rho(\psi) .
\end{equation}

A {\it classical statistical model} is a pair $M=(S, V).$ We recall
that classical statistical mechanics on the phase space
$\Omega_{2n}= {\bf R}^n\times {\bf R}^n$ gives an example of a
classical statistical model. But we shall not be interested in this
example in our further considerations. We shall develop  a classical
statistical model with {\it an infinite-dimensional phase-space.}

In real Hilbert space $H$ a quantum statistical model is described
in the following way (see Dirac-von Neumann [2], [4] for the
conventional complex model): a) physical observables are represented
by operators $A: H \to H$ belonging to the class of continuous
self-adjoint operators ${\cal L}_s \equiv {\cal L}_s (H);$ b)
statistical states are represented by von Neumann density operators,
see [4] (the class of such operators is denoted by  ${\cal D} \equiv
{\cal D} (H));$ d) the average of a physical observable (which is
represented by the operator
 $A \in {\cal L}_s (H))$ with respect to a statistical state (which is represented
  by the density operator $D \in {\cal D} (H))$ is given by von Neumann's
formula [4]:
\begin{equation}
\label{AV1} <A >_D \equiv \rm{Tr}\; DA
\end{equation}
The {\it quantum statistical model} is the pair $N_{\rm{quant}}
=({\cal D}, {\cal L}_s).$

\subsection{Asymptotic equality of classical and quantum averages and amplification of classical variables}

We are looking for a classical statistical model $M=(S, V)$ which
will give ``dequantization'' of the quantum model $N_{\rm{quant}}
=({\cal D}, {\cal L}_s).$ Here the meaning of ``dequantization''
should be specified. In fact, all ``NO-GO'' theorems (e.g., von
Neumann, Kochen-Specker, Bell,...) can be interpreted as theorems
about impossibility of various dequantization procedures. Therefore
we should define the procedure of dequantization in such a way that
there will be no contradiction with known ``NO-GO'' theorems, but
our dequantization procedure still will be natural from the physical
viewpoint. We define (asymptotic)  dequantization as a family
$M^\alpha=(S^\alpha, V)$ of classical statistical models depending
on small parameter $\alpha.$  There  should exist maps
$T:S^\alpha\to {\cal D}$ and $T: V \to  {\cal L}_s$ such that: a)
both maps are {\it surjections} (so all quantum objects are covered
by classical); b) the map $T: V \to {\cal L}_s$ is linear; c) the
map $T:S\to {\cal D}$ is injection (there is one-to one
correspondence between classical and quantum statistical states); d)
classical and quantum averages are coupled through the following
asymptotic equality:
 \begin{equation}
\label{AQ} < f >_\rho = \alpha <T(f)>_{T(\rho)} + o(\alpha), \; \;
\alpha \to 0
\end{equation}
(here $<T(f)>_{T(\rho)}$ is the  quantum average).
 In mathematical models this equality has the form:
\begin{equation}
\label{AQ1} \int_\Omega f(\psi) d \rho(\psi)=  \alpha \; \Tr \; D A
+ o(\alpha), \; \; A=T(f), D= T(\rho).
\end{equation}
This equality can be interpreted in the following way. Let $f(\psi)$
be a classical physical variable (describing properties of
microsystems - classical fields having very small magnitude
$\alpha).$  We define its {\it amplification} by:
 \begin{equation}
\label{AMP} f_\alpha (\psi) =\frac{1}{\alpha} f(\psi)
\end{equation}
(so any micro effect is amplified in $\frac{1}{\alpha}$-times). Then
we have:
\begin{equation} \label{AQ4} < f_\alpha >_\rho =
<T(f)>_{T(\rho)} + o(1), \; \; \alpha \to 0,
\end{equation}
or
\begin{equation} \label{AQ5}
\int_\Omega f_\alpha(\psi) d \rho(\psi)=   \Tr \; D A + o(1), \; \;
A=T(f), D= T(\rho).
\end{equation}
Thus: {\it Quantum average $\approx$ Classical average of the
$\frac{1}{\alpha}$-amplification.} Hence: {\it QM is a mathematical
formalism describing a statistical approximation of amplification of
micro effects.}

We see that for physical variables/quantum observables and classical
and quantum statistical states the dequantization  maps have
different features. The map $T: V\to  {\cal L}_s$ is not injective.
Different classical physical variables $f_1$ and $f_2$ can be mapped
into one quantum observable $A.$ This is not surprising. Such a
viewpoint on the relation between classical variables and quantum
observables was already presented by J. Bell, see [12]. In
principle, experimenter could not distinguish classical (``ontic'')
variables by his measurement devices. In contrast, the map
$T:S^\alpha\to {\cal D}$ is injection. Here we suppose that quantum
statistical states represent uniquely  (``ontic'') classical
statistical states.

The crucial difference with dequantizations considered in known
``NO-GO'' theorems is that in our case classical and quantum
averages are equal only asymptotically and that a classical variable
$f$ and the corresponding quantum observable $A=T(f)$ can have
different ranges of values.

\subsection{Asymptotic Gaussian analysis}

Let us consider a classical statistical model in that the state
space $\Omega= H$ (in physical applications $H=L_2({\bf R}^3)$ is
the space of classical fields on  ${\bf R}^3)$ and the space of
statistical states consists of Gaussian measures with zero mean
value and dispersion
\begin{equation}
\label{DS}\sigma^2 (\rho)= \int_\Omega \Vert \psi \Vert^2 d
\rho(\psi)= \alpha,
\end{equation}where $ \alpha> 0$ is a small real parameter. Denote such a class of Gaussian measures by the
symbol $S_G^\alpha(\Omega).$ For $\rho \in S_G^\alpha(\Omega),$ we
have
$\rm{Tr} \; \rm{cov} \; \rho =  \alpha.$
We remark that any linear transformation (in particular, scaling)
preserves the class of Gaussian measures. Let us make the change of
variables (scaling):
\begin{equation}
\label{LHT}  \psi \to \frac{ \psi}{\sqrt{\alpha}} .
\end{equation}
(we emphasize that this is a scaling not in the physical space ${\bf
R}^3,$ but in the space of fields on it). To find the covariation
operator $D$ of the image  $\rho_D$  of the Gaussian measure
$\rho_B,$ we compute its Fourier transform: $\tilde\rho_D
(\xi)=\int_\Omega e^{i(\xi, y)} d\rho_D(y)= \int_\Omega e^{i(\xi,
\frac{\psi}{\sqrt{\alpha}})} d\rho_B (\psi)= e^{-\frac{1}{2\alpha}(B
\xi, \xi)}.$ Thus
\begin{equation}
\label{LL2} D=\frac{B}{\alpha}=\frac{{\rm cov} \rho}{\alpha}.
\end{equation}
We shall use this formula later. We remark that by definition:
$$
<f>_{\rho_B} = \int_\Omega f( \psi) d\rho_B ( \psi)= \int_\Omega
f(\sqrt{\alpha}  \psi) d\rho_D (\psi).
$$
To make our further
considerations mathematically rigorous, we should attract the theory
of analytic functions $f: \Omega^{{\bf C}}\to {\bf C}.$ Here
$\Omega^{{\bf C}}= \Omega \oplus i \Omega$ is the complexification
of the real Hilbert  space $\Omega.$

Let $b_n:\Omega^{{\bf C}}\times ...\times \Omega^{{\bf C}} \to {\bf
C}$ be a continuous $n$-linear symmetric form. We define its norm by
$ \Vert b_n \Vert =\sup_{\Vert \psi \Vert \leq 1} \vert b_n
(\psi,...,\psi)\vert. $ Thus
\begin{equation}
\label{INT} \vert b_n (\psi,...,\psi)\vert \leq \Vert b_n \Vert
\Vert \psi \Vert^n
\end{equation}
Let us consider the space  analytic functions of the {\it
exponential growth:} \begin{equation} \label{ZO2} \vert f(\psi)\vert
\leq a e^{b\Vert  \psi \Vert}, \psi \in \Omega^{{\bf C}},
\end{equation}
see, e.g., [60]. Here constants depend on $f: \; a=a_f, b=b_f.$

\medskip

{\bf Lemma 3.1}  {\it The space of analytic functions of the
exponential growth coincides with the space of analytic functions
such that:
\begin{equation}
\label{ZO1} \Vert f^{(n)} (0) \Vert \leq  c\;  r^n, \; n=0,1, 2,...
\end{equation}
Here constants $c=c_f$ and $r=r_f$ depend on the function $f.$}

{\bf Proof.} A). Let $f$ have the exponential growth. For any $\psi
\in \Omega^{{\bf C}},$ we consider the function of the complex
variable $z\in {\bf C}: g_\psi(z)= f(z\psi).$ By the Cauchy integral
formula for $g_\psi(z)$ we have: $g_\psi^{(n)}(0)=\frac{n!}{2\pi i}
\int_{\vert z\vert=R} g_\psi(z) z^{-(n+1)} dz,$ where at the moment
$R>0$ is a free parameter.  Thus: $\vert g_\psi^{(n)}(0) \vert \leq
n! R^{-n}\sup_{0\leq \theta \leq 2 \pi} \vert f(R e^{i \theta}
\psi)\vert\leq a_f n! R^{-n} e^{b_f R \Vert \psi \Vert}.$ By
choosing $R=n$ and observing that $g_\psi^{(n)}(0)=
f^{(n)}(0)(\psi,..., \psi)$ we obtain:
$$
\Vert f^{(n)}(0) \Vert \leq a_f^\prime e^{-n} n^{1/2} e^{b_f n}.
$$
Thus the derivatives of $f$ satisfy the inequalities (\ref{ZO1})
with $r_f=e^{b_f}.$

B). Let now derivatives of $f$ satisfy the inequalities (\ref{ZO1}).
Then by the inequalities (\ref{INT}) we have $\vert f(\psi) \vert
\leq \sum_{n=0}^\infty \Vert f^{(n)}(0) \Vert \Vert \psi
\Vert^n/n!\leq c_f \sum_{n=0}^\infty (r_f \Vert \Vert \psi \Vert)^n/
n!\leq c_f e^{r_f \Vert \psi \Vert.}$ Thus $f$ has the exponential
growth with $b_f=r_f.$

\medskip

We denote by the symbol ${\cal V}(\Omega)$ the following space of
functions $f: \Omega \to {\bf R}.$ Each $f \in {\cal V}(\Omega)$
takes the value zero at the point  $\psi=0$ and it can be extended
to the analytic function $f: \Omega^{{\bf C}} \to {\bf C}$ having
the exponential growth.

{\bf Example 3.1.} In particular, any polynomial on $\Omega$ belongs
to the space ${\cal V}(\Omega).$ For example, let $A_1,..., A_N$ be
continuous linear operators. Then function $f(\psi)=\sum_{n=1}^N
(A_n\psi,\psi)^n$ belongs to the space ${\cal V}(\Omega).$

Any function $f\in {\cal V}(\Omega)$ is integrable with respect to
any Gaussian measure on $\Omega,$ see [55].  Let us consider the
family of the classical statistical models
$$
M^\alpha=(S_G^\alpha(\Omega), {\cal V}(\Omega)).
$$

Let a variable $f \in {\cal V}(\Omega)$ and let a statistical state
$\rho_B\in S_G^\alpha(\Omega).$ Our further aim is to find an
asymptotic expansion of the (classical) average
$<f>_{\rho_B}=\int_\Omega f( \psi) d\rho_B( \psi)$ with respect to
the small parameter $\alpha.$

\medskip

{\bf Lemma 3.2.} {\it Let $f \in {\cal V}(\Omega)$ and let $\rho \in
S_G^\alpha(\Omega).$ Then the following asymptotic equality holds:
\begin{equation}
\label{ANN3} <f>_\rho =  \frac{\alpha}{2} \; \rm{Tr}\; D \;
f^{\prime \prime}(0) + o(\alpha), \; \alpha \to 0,
\end{equation}
where the operator $D$ is given by (\ref{LL2}). Here
\begin{equation}
\label{OL} o(\alpha) = \alpha^2 R(\alpha, f, \rho),
\end{equation}
where $\vert R(\alpha,f,\rho)\vert \leq c_f\int_\Omega  e^{r_f \Vert
\psi \Vert}d\rho_D (\psi).$

}

{\bf Proof.} In the Gaussian integral $\int_\Omega f( \psi) d\rho(
\psi)$ we make the scaling (\ref{LHT}):
\begin{equation}
\label{ANN1} <f>_{\rho}= \int_\Omega f(\sqrt{\alpha}  \psi) d\rho_D
( \psi)= \frac{\alpha}{2} \int_\Omega (f^{\prime \prime}(0)\psi,
\psi) \; d\rho_D(\psi) + \alpha^2 R(\alpha,f,\rho),
\end{equation}
where
$$
R(\alpha,f,\rho)= \int_\Omega g(\alpha,f; \psi) d\rho_D (\psi),
g(\alpha,f; \psi)= \sum_{n=4}^\infty \frac{\alpha^{n/2-2}}{n!}
 f^{(n)}(0)( \psi, ..., \psi).
$$
 We pay attention that
$$ \int_\Omega (f^\prime(0),  \psi) d\rho_D(
\psi)=0,\; \;\; \; \int_\Omega
f^{\prime\prime\prime}(0)(\psi,\psi,\psi) d\rho_D( \psi)=0,
$$
because the mean value of $\rho$  (and, hence, of $\rho_D)$ is equal
to zero. Since $\rho\in S_G^\alpha(\Omega),$ we have $\rm{Tr} \; D =
1.$

The change of variables in (\ref{ANN1}) can be considered as scaling
of the magnitude of statistical  (Gaussian) fluctuations. Negligibly
small random  fluctuations $\sigma (\rho)= \sqrt{\alpha}$ (where
$\alpha$ is a small parameter) are considered in the new scale as
standard normal fluctuations. If we use the language of probability
theory and consider a Gaussian random variables $\xi(\lambda),$ then
the transformation (\ref{LHT}) is nothing else than the standard
normalization of this random variable (which is used, for example,
in the central limit theorem): $\eta(\lambda)= \frac{\xi(\lambda) -
E \xi}{\sqrt{E(\xi(\lambda) - E \xi)^2}}$ (in our case $E \xi=0).$

We now estimate the rest term $R(\alpha,f,\rho).$ By using the
inequality (\ref{ZO1}) we have for $\alpha \leq 1:$
$$
\vert g(\alpha,f; \psi)\vert  = \sum_{n=4}^\infty  \frac{\Vert
f^{(n)} (0) \Vert \Vert \psi \Vert^n}{n!} \leq c_f \sum_{n=4}^\infty
\frac{r_f^n  \Vert \psi \Vert^n}{n!}= C_f e^{r_f \Vert \psi \Vert}.
$$
Thus: $ \vert R(\alpha,f,\rho)\vert \leq c_f\int_\Omega  e^{r_f
\Vert \psi \Vert}d\rho_D (\psi). $ We obtain:
\begin{equation}
\label{ANN2} <f>_\rho=  \frac{\alpha}{2} \int_\Omega (f^{\prime
\prime}(0)\psi, \psi) \; d\rho_D(\psi) + o(\alpha), \; \alpha \to 0.
\end{equation}
By using the equality (\ref{QI}) we finally come the asymptotic
equality (\ref{ANN3}).

\medskip

We see that the classical average (computed in the model
$M^\alpha=(S_G^\alpha(\Omega),{\cal V}(\Omega))$ by using the
measure-theoretic approach) is coupled through (\ref{ANN3}) to the
quantum average (computed in the model $N_{\rm{quant}} =({\cal
D}(\Omega),$ ${\cal L}_{{\rm s}}(\Omega))$ by the von Neumann
trace-formula).

The equality (\ref{ANN3}) can be used as the motivation for defining
the following classical $\to$ quantum map $T$ from the classical
statistical model $M^\alpha=(S_G^\alpha, {\cal V})$ onto the quantum
statistical model $N_{\rm{quant}}=({\cal D}, {\cal L}_{{\rm s}}):$
\begin{equation}
\label{Q20} T: S_G^\alpha(\Omega) \to {\cal D}(\Omega), \; \;
D=T(\rho)=\frac{\rm{cov} \; \rho}{\alpha}
\end{equation}
(the Gaussian measure $\rho$ is represented by the density matrix
$D$ which is equal to the covariation operator of this measure
normalized by  $\alpha$);
\begin{equation}
\label{Q30}T: {\cal V}(\Omega) \to {\cal L}_{{\rm s}}(\Omega), \; \;
A_{\rm quant}= T(f)= \frac{1}{2} f^{\prime\prime}(0).
\end{equation}
Our previous considerations can be presented as

\medskip

{\bf Theorem 3.1.} {\it The one parametric family of classical
statistical models $M^\alpha=(S_G^\alpha, {\cal V})$ provides
dequantization of the quantum model $N_{\rm{quant}}=({\cal D}, {\cal
L}_s)$ through the pair of maps (\ref{Q20}) and (\ref{Q30}). The
classical and quantum averages are coupled by the asymptotic
equality (\ref{ANN3}).}

\section{Gaussian underground for pure states} In quantum mechanics  a pure quantum
state is given by a normalized vector $\Psi \in H: \ \Vert \Psi
\Vert=1.$ In our model such a state is not pure at all (in the sense
that such a vector $\Psi$ does not provide a description of an
individual system). Such a normalized vector $\Psi$ is the label of
a Gaussian statistical mixture. The corresponding  quantum
statistical state is represented by the density operator:
$D_\Psi=\Psi \otimes \Psi.$ In particular, the von Neumann's
trace-formula for expectation has the form: $\rm{Tr}\;  D_\Psi A=(A
\Psi, \Psi).$ Let us consider the correspondence map $T$ for
statistical states for the classical statistical model
$M^\alpha=(S_G^\alpha, {\cal V}),$ see (\ref{Q20}). A pure quantum
state $\Psi$ (i.e., the state with the density operator $D_\Psi)$ is
the image of the Gaussian statistical mixture $\rho_\Psi$ of states
$\psi \in H.$ We use the capital $\Psi$ to denote a quantum pure
state. This is just the special system of labeling of the Gaussian
measure $\rho_\Psi$ by the normalized vector $\Psi$ of Hilbert
space. Points of the sample space on that this measure is defined we
denote by the low $\psi.$ The measure $\rho_\Psi$ has the
covariation operator $B_\Psi= \alpha D_\Psi.$ This means that the
measure $\rho_\Psi$ is concentrated on the one-dimensional subspace
$ H_\Psi=\{x \in H: x=s\Psi, s \in {\bf R}\}. $ This is
one-dimensional Gaussian distribution.

\section{Pure states as one-dimensional projections of
spatial white-noise} In section 4 we showed that so called pure
states of quantum mechanics have the natural classical statistical
interpretation as Gaussian measures concentrated on one-dimensional
subspaces of the Hilbert space $H.$ On the other hand, it is well
known that {\it any Gaussian measure on $H$ is determined by its
one-dimensional projections.} To determine a Gaussian random
variable $\xi(\omega) \in H,$ it is sufficient to determine all its
one-dimensional projections: $\xi_\Psi (\omega)=(\Psi, \xi(\omega)),
\Psi \in H.$ The covariation operator $B$ of $\xi$ (having the zero
mean value) is defined by $(B\Psi, \Psi)=E \xi_\Psi^2.$ We are
interested in the following problem:

\medskip

{\it Is it possible to construct a Gaussian distribution on $H$ such
that its one-dimensional projections will give us all pure quantum
states, $\Psi\in H, \Vert \Psi \Vert=1?$}

\medskip

We recall that in our approach a pure quantum state $\Psi$ is just
the label for a Gaussian  random variable $\xi_\Psi$ such that  $E
\xi_\Psi^2=\alpha ||\Psi||^2=\alpha.$  Thus the answer to our
question is positive and pure quantum states can be considered as
one-dimensional projections of the $\sqrt{\alpha}$-scaling of the
standard Gaussian distribution on $H.$ The standard Gaussian
distribution $\mu$ on $H$ (so the average of $\mu$ is equal to zero
and ${\rm cov} \; \mu=I,$ where $I$ is the unit operator) is nothing
else than the {\it white noise} on ${\bf R}^3$ (if one chooses
$H=L_2 ({\bf R}^3)$), see [56]--[59] for details. Thus pure quantum
states are simply one-dimensional projections of the {\it spatial
white noise.} It is well known, see, e.g., [56]--[59], that the
$\mu$ {\it is not $\sigma$-additive on the $\sigma$-field of Borel
subsets of $H$.}

To escape mathematical difficulties and concentrate on the
dequantization of quantum mechanics, we start with consideration of
the finite-dimensional case.

\subsection{The finite-dimensional case}
We consider the family of Gaussian random variables $\xi_\Psi, \Psi
\in {\bf R}^n,$  $E \xi_\Psi=0, E \xi_\Psi^2=\alpha||\Psi||^2.$ This
family can be realized as $\xi_\Psi(\omega)=(\Psi, \xi (\omega))$
where $\xi (\omega)=\sqrt{\alpha} \eta (\omega)$ and $\eta (\omega)
\in {\bf R}^n$ is standard Gaussian random variable (so $E \eta=0,
{\rm cov}\; \eta=I).$  For any $\Psi \in {\bf R}^n,$ we define the
projection $P_\Psi$ to this vector: $P_\Psi (k)=(\Psi, k) \Psi.$

Denote by the symbol ${\cal V}({\bf R}^n)$ the class of functions
$f:{\bf R}^n \to {\bf R}$ such that $f(0)=0$ and $f$ can be
continued analytically onto ${\bf C}^n$ and this continuation $f(z)$
has the exponential growth.

{\bf Proposition 5.1} {\it Let $f \in {\cal V}({\bf R}^n).$  Then we
have:}
\begin{equation}
\label{ZX} E f(P_\Psi \xi (\omega))=\frac{\alpha||\Psi||^2}{2}
(f^{\prime\prime}(0) \Psi, \Psi) + o(\alpha), \alpha \to 0.
\end{equation}

{\bf Proof.} By using the Taylor expansion of $f$ we obtain:
$$
E f(P_\Psi \xi (\omega))=\frac{1}{2} E (\Psi, \xi(\omega))^2
(f^{\prime\prime}(0) \Psi, \Psi) + o(\alpha), \alpha \to 0.
$$
By setting into this asymptotic equality the dispersion of the
random variable $P_\Psi \xi (\omega)$ we obtain (\ref{ZX}).

\medskip

If $||\Psi||=1$ (a pure quantum state), then we get:
\begin{equation}
\label{TE} E f(P_\Psi \xi(\omega))=\frac{\alpha}{2}
(f^{\prime\prime}(0) \Psi, \Psi) + o(\alpha), \alpha \to 0.
\end{equation}
Here $A=f^{\prime\prime}(0)$ is a symmetric linear operator. We
``quantize'' the classical variable $f(x), x \in {\bf R}^n,$ by
mapping it to the operator $A=\frac{1}{2} f^{\prime\prime} (0),$ see
Theorem 3.1. The Gaussian random variable $\xi_\Psi, ||\Psi||=1$ is
``quantize'' by mapping it into the pure quantum state $\Psi.$

{\bf Theorem 5.1.} {\it There exists a Kolmogorov probability space
such that all pure quantum states can be represented by Gaussian
random variables on this space. The correspondence $\Psi \to
\xi_\Psi (\omega)$ is linear:
\begin{equation}
\label{ZX1} \lambda_1 \Psi_1 + \lambda_2 \Psi_2 \to \lambda_1
\xi_{\Psi_1}(\omega) + \lambda_2 \xi_{\Psi_2}(\omega) ,
\end{equation}
where $\lambda_1, \lambda_2 \in {\bf R}.$}

{\bf Proof.} We choose $\Omega={\bf R}^n$ as the space of elementary
events, the $\sigma$-field of Borel subsets is the space of events
and the standard Gaussian measure $\mu$ as the probability measure.
Then for $\Psi \to \xi_\Psi(\omega)=\sqrt{\alpha}(\Psi, \omega),
\omega \in {\bf R}^n,$ we have: $\lambda_1 \xi_{\Psi_1}(\omega) +
\lambda_2 \xi_{\Psi_2}(\omega)=\lambda_1 (\Psi_1, \omega) +
\lambda_2 (\Psi_2, \omega)=(\lambda_1 \Psi_1 + \lambda_2 \Psi_2,
\omega)=\xi_{\lambda_1 \Psi_1 + \lambda_2 \Psi_2}(\omega).$

\medskip

This theorem is rather surprising from the common viewpoint (by that
essentially nonclassical probabilistic features of quantum states
are consequences of the  {\it non-Kolmogorovian structure} of the
quantum probabilistic model).

We pay attention that physical variables $ \xi_\Psi (\omega)=P_\Psi
\xi(\omega), \Psi \in {\bf R}^n,$ (one-dimensional projections of
the scaling $\xi(\omega)$ of the standard Gaussian random variable
$\eta(\omega) \in {\bf R}^n)$ {\it cannot be mapped onto nontrivial
quantum observables.} Prequantum classical physical variables
$\xi_\Psi(\omega)=(\Psi, \omega)$ are linear functionals of
$\omega.$ Therefore $T(\xi_\Psi)=\xi_\Psi^{\prime\prime}(0)=0.$
Nevertheless, quantum mechanics contains images of $\xi_\Psi$ given
by quantum states $\Psi, $ but only for $\Psi$ with $||\Psi||=1!$

We call $\xi(\omega)$ a {\it background random field.} All pure
states could be extracted from the the background random field by
projecting it to one dimensional subspaces. PCSFT explains the
origin of the scalar product on the set of pure quantum states. We
consider the $1/\alpha$-amplification of the covariation of two
Gaussian (prequantum) random variables $\xi_{\Psi_1}(\omega)$ and
$\xi_{\Psi_2}(\omega).$ We have:
\begin{equation}
\label{ZXY} \frac{1}{\alpha} E \xi_{\Psi_1}(\omega)
\xi_{\Psi_2}(\omega)=(\Psi_1, \Psi_2).
\end{equation}

{\bf Conclusion.} {\it The Hilbert space structure of quantum
mechanics is induced by the (prequantum)  Gaussian random field (the
background field $\xi(\omega))$ through the $\alpha\to 0$
asymptotic.}

At the moment we proved this only in the finite-dimensional case. In
section 5.2 we shall
 do this in the infinite-dimensional case. Finally, we pay attention to the fact that,
 for quadratic physical variables $f(x)=\frac{1}{2} (Ax, x),$ where $A:H \to H$ is a
 symmetric operator, the asymptotic equality (\ref{TE}) is reduced to the precise equality
  of averages. By considering directly the standard Gaussian random variable $\eta(\omega)$
  (instead of the background random field $\xi(\omega)=\sqrt{\alpha} \eta (\omega)$)
  we come to the following classical probabilistic representation of the quantum average:
$E f(P_\Psi \eta (\omega))=(A \Psi, \Psi)\equiv <A>_\Psi.$

\subsection{Prequantum white noise field}
To repeat consideration of section 5.1 for the infinite-dimensional
case, we consider measures on the so called rigged Hilbert spaces.
 We apply some rather abstract mathematical
constructions. However, finally we shall consider a simple concrete
example which will be then used as the basis of our prequantum
classical statistical model.

Let $\Omega$ be a nuclear Frechet\footnote{so complete metrizable
and locally convex} topological linear space and $\Omega^\prime$ its
dual space. Suppose that $\Omega$ is densely and continuously
embedded into a Hilbert space $H,$ so $\Omega \subset H.$ Thus the
dual space $H^\prime$ is densely embedded into $\Omega^\prime.$ By
identifying $H$ and $H^\prime$ we obtain the rigged Hilbert space:
\begin{equation}
\label{GT} \Omega\subset H \subset  \Omega^\prime
\end{equation}
In our final application we shall set $\Omega={\cal S}({\bf R}^3).$
This is the space of Schwartz test functions on ${\bf R}^3.$ Here
$\Omega^\prime={\cal S}^\prime({\bf R}^3)$ is the space of Schwartz
distributions. In this case we choose $H=L_2({\bf R}^3)$ and we
shall consider the rigged Hilbert space:
\begin{equation}
\label{GN} {\cal S}({\bf R}^3) \subset L_2 ({\bf R}^3) \subset {\cal
S}^\prime({\bf R}^3)
\end{equation}
Readers who are not so much interested in general theory of
topological linear spaces can consider this rigged Hilbert space
throughout this section.

A Gaussian measure $\rho$ on $\Omega^\prime$ is determined by its
characteristic functional (Fourier transform) $\tilde \rho$ which is
defined on $\Omega:$ $\tilde \rho (\Psi)=e^{-\frac{1}{2}b(\Psi,
\Psi)},$ where $b: \Omega \times \Omega \to {\bf R}$ is a continuous
positively defined quadratic form. By the well known theorem of
Minlos-Sazonov, see e.g., [], $\rho$ is $\sigma$-additive on
$\Omega^\prime$ and its covariation functional is equal to $b.$ Here
$b(\Psi_1, \Psi_2)=\int_{\Omega^\prime} (\phi, \Psi_1) (\phi,
\Psi_2) d\rho(\phi),$ where $\Psi_1, \Psi_2 \in \Omega.$ This
functional defines the covariation operator $B: \Omega \to
\Omega^\prime$ by $(B \Psi_1, \Psi_2)= b(\Psi_1, \Psi_2).$ This
operator is self-adjoint in the following sense. The dual operator
$B^\prime: \Omega^{\prime\prime} \to \Omega^\prime.$ But, since the
topological linear space $\Omega$ is a nuclear Frechet space, it is
{\it reflexive.}  Hence, $\Omega^{\prime\prime} = \Omega.$ Thus the
operator $B^\prime: \Omega \to \Omega^\prime.$ Thus it is meaningful
to speak about self-adjoint operators in this framework (by
extending the ordinary theory of self-adjoint operators in Hilbert
space). We also pay attention to the fact that the covariation
operator $B$ is positively defined.

Let us consider the standard Gaussian distribution $\mu $ on $H$
that is defined by its covariation functional:
$$b(\Psi_1, \Psi_2)=(\Psi_1, \Psi_2).$$
The corresponding covariation operator $B=I: \Omega \to
\Omega^\prime$ is the canonical embedding operator. Since the
embedding $\Omega \subset H$ is continuous, $b: \Omega \times \Omega
\to {\bf R}$ is continuous and, hence, the measure $\mu$ is
$\sigma$-additive on $\Omega^\prime.$ Therefore there is well
defined the corresponding Gaussian random variable $\eta(\phi) \in
\Omega^\prime.$

In the case of the rigged Hilbert space (\ref{GN}) the Gaussian
random field $\eta(\phi)\in {\cal S}^\prime({\bf R}^3)$ is nothing
else than the {\it spatial white noise.} We extend this terminology
and we shall call $\eta(\phi)$ {\it white noise} even in the
abstract framework. Let us consider  $\sqrt{\alpha}$-scaling of
white noise
$$\xi(\phi)=\sqrt{\alpha}\eta(\phi)$$ and its one-dimensional projections:
$\xi_\Psi(\phi)=(\xi(\phi), \Psi), \Psi \in \Omega.$ We have $E
\xi_\Psi=0, E \xi_\Psi^2=\alpha||\Psi||^2.$ The $\xi(\phi)$ is the
{\it background field} in our prequantum model (PCSFT).

For any $\Psi \in \Omega,$ we consider the one-dimensional projector
$P_\Psi(\phi)=(\phi, \Psi) \Psi, \phi \in \Omega^\prime, $ and the
$\Omega$-valued random variable $P_\Psi \xi(\phi)=\xi_\Psi(\phi)
\Psi.$ If $||\Psi||=1,$ then the $T$-image of the corresponding
Gaussian distribution $\rho_\Psi$ is nothing else than the pure
state $\Psi.$

This correspondence can be extended form the space $\Omega$  to the
Hilbert space $H.$ If $\Psi \in H,$ then $\xi_\Psi(\phi)=(\phi,
\Psi)$ is also well defined, but is is not a continuous linear
functional on the space $\Omega^\prime.$ The $\xi_\Psi(\phi)$ is
defined as an element of the space of square integrable functionals
of the white noise: $\xi_\Psi \in L_2(\Omega^\prime, d \mu).$ To
define $\xi_\Psi,$ we approximate $\Psi \in H$ by elements $\Psi_n$
of $\Omega, \Psi_n \to \Psi$ in $H$ (we recall that $\Omega$ is
dense in $H).$
 Then
$\xi_\Psi=\lim_{n \to \infty} \xi_{\Psi_n}$ in $L_2 (\Omega^\prime,
d \mu).$

\medskip

{\bf Lemma 5.1.} {\it Let $f: H \to {\bf R}$ be a polynomial and let
$f(0)=0.$ Then, for any $\Psi \in H,$  the asymptotic equality
(\ref{ZX}) holds.}

{\bf Proof.}  Here the main difference from consideration in section
3 is that the measure $\mu$ is not concentrated on Hilbert space $H$
on which the function $f$ is defined (and continuous). Therefore
even the exponential growth of $f$ on $H$ would not help so much,
because $\int_{\Omega^\prime} e^{a\Vert \phi \Vert} d \mu(\phi) =
\infty$ (since even $\int_{\Omega^\prime} \Vert \phi \Vert d
\mu(\phi) = \infty).$ We have for a polynomial $f$:
$$
E f(P_\Psi \xi(\phi))= \sum_{k=1}^N \frac{f^{(2k)}(0)(\Psi,...,
\Psi)}{2k!} E \xi_\Psi^{2k}(\phi) $$
$$
= \sum_{k=1}^N \frac{f^{(2k)}(0)(\Psi,..., \Psi)}{(2k)!}
\frac{\alpha^{2k} \Vert \Psi \Vert^{2k} (2k)!}{2^k k!}.
$$
Since the sum is finite and derivatives of $f$ are continuous forms
on $H,$ we obtain (\ref{ZX}).

\medskip

We  ``quantize'' $f(u)$ by mapping it into
$\frac{f^{\prime\prime}(0)}{2}.$ For quadratic functionals
$f(u)=\frac{1}{2}(Au, u), A \in {\cal L}_s(H),$ we have the precise
equality and we can directly use the average with respect to the
canonical Gaussian random variable $\eta(\omega).$ Here
$$
E f(P_\Psi \eta (\omega))=\frac{1}{2}(f^{\prime\prime}(0) \Psi,
\Psi).
$$

\section{Unbounded operators}

In this section we shall use theory of Gaussian measures on
topological vector spaces, see, e.g., Smolyanov and Fomin [66] for
detail.

Let $f: \Omega \to {\bf R}$ be a smooth function. Then, at any point
$\psi_0 \in \Omega,$ $f^{\prime\prime}(\psi_0): \Omega \to
\Omega^\prime.$ Therefore $f^{\prime\prime}(0)$ is in general
unbounded operator in $H$.

Moreover,  in this way (i.e., starting with PCSFT) we obtain the
class of linear operators (quantum observables) that is even
essentially larger than in the conventional quantum formalism. In
general, $A=f^{\prime\prime}(0)$ maps $\Omega$ not into $H$, but
into $\Omega^\prime.$

{\bf Example 6.1.}  Let us consider the rigged Hilbert space
(\ref{GN}). We consider the map $f:{\cal S}({\bf R}^3)\to {\bf R}$
determined by a fixed point $x_0 \in {\bf R}^3:$
$$f(\psi)=\frac{1}{2} \psi^2(x_0).$$
(For example, the classical field $\psi(x)=e^{-x^2}$ is mapped into
the real number $e^{-x_0^2}).$ Then $(f^{\prime\prime}(0)\psi_1,
\psi_2)=\psi_1(x_0) \psi_2 (x_0).$ Thus
$$A \psi(x)=
\frac{1}{2}f^{\prime\prime}(0) \psi(x)=\frac{1}{2} \delta(x-x_0)
\psi(x)$$ is the operator of multiplication by the $\delta$-function
$\delta(x-x_0).$ Hence
$$f^{\prime\prime}(0)({\cal S}({\bf R}^3))
\not\subset L_2({\bf R}^3).
$$
For any $\Psi \in {\cal S}({\bf R}^3),$  we have
$$
E f(P_\Psi \eta (\omega))=\frac{1}{2}\Psi^2(x_0)=(A\Psi, \Psi)\equiv
<A>_\Psi.
$$
However, in general for $\Psi \in L_2 ({\bf R}^3)$ the average
$<A>_\Psi$ is not well defined.

\medskip

We can consider not only pure states, but general density operators.
Let us now consider a Gaussian measure $\rho \in S_G^\alpha (H)$
which has the support on the space $\Omega.$ Thus $\rho$ can be
considered as a measure on $\Omega.$ For such a measure $\rho$ its
covariation operator $B: \Omega^\prime \to \Omega$ and its Fourier
transform $\tilde{\rho}$ is defined on $\Omega^\prime.$ We denote
this class of statistical states by the symbol $S_G^\alpha
(\Omega).$ We remark that $S_G^\alpha (\Omega) \subset S_G^\alpha
(H).$

Let $E$ be a complex locally convex topological linear space. We
recall that the topology of $E$ can be determined by a system of
semi-norms (the notion of a semi-norm $p$ generalizes the notion of
a norm $\Vert \cdot \Vert;$ the only difference is that $p(\psi)$
can be equal to zero even for a nonzero vector $\psi).$ Let $b_n: E
\times...\times E  \to {\bf C}$ be a continuous $n$-linear symmetric
form. There exits a continuous semi-norm $p$ on $E$   such that
$$
\Vert b_n  \Vert_p =\sup_{p(\psi) \leq 1} \vert b_n
(\psi,...,\psi)\vert < \infty
$$
(here $p\equiv p_{b_n}).$ Thus
\begin{equation}
\label{INTZ} \vert b_n (\psi,...,\psi)\vert \leq \Vert b_n \Vert \;
p^n(\psi)
\end{equation}
An analytic function, see, e.g., [60] for details, $f:E\to {\bf C}$
has the exponential growth if there exits a continuous semi-norm $p$
on $E$ on  such that:
\begin{equation}
\label{WS} \vert f(\psi)\vert \leq a e^{b p(\psi)}, \psi \in E.
\end{equation}
Here the constants and the semi-norm depend on $f: a\equiv a_f,
b\equiv b_f, p\equiv p_f.$

\medskip

{\bf Lemma 6.1.}  {\it The space of analytic functions of the
exponential growth coincides with the space of analytic functions
such that there exists a continuous semi-norm $p=p_f$:
\begin{equation}
\label{ZO1Z} \Vert f^{(n)} (0) \Vert_p \leq  c\;  r^n, \; n=0,1,
2,...
\end{equation}
Here constants $c=c_f$ and $r=r_f$ depend on the function $f.$}

{\bf Proof.} A). Let $f$ have the exponential growth. For any $\psi
\in E,$ we consider the function of the complex variable $z\in {\bf
C}: g_\psi(z)= f(z\psi).$ As in Lemma 5.1, we have: $\vert
g_\psi^{(n)}(0) \vert \leq n! R^{-n}\sup_{0\leq \theta \leq 2 \pi}
\vert f(R e^{i \theta} \psi)\vert\leq a_f n! R^{-n} e^{b_f R
p(\psi)}.$ We obtain:
$$
\Vert f^{(n)}(0) \Vert_p \leq a_f^\prime e^{-n} n^{1/2} e^{b_f n}.
$$
Thus the derivatives of $f$ satisfy the inequalities (\ref{ZO1Z})
with $r_f=e^{b_f}.$

B). Let now derivatives of $f$ satisfy the inequalities (\ref{ZO1Z})
for some continuous semi-norm  $p.$ Then by the inequalities
(\ref{INTZ}) we have
$$
\vert f(\psi) \vert \leq \sum_{n=0}^\infty \Vert f^{(n)}(0) \Vert_p
p^n(\psi)/n! \leq c_f e^{r_f  p(\psi)}.
$$
Thus $f$ has the exponential growth with $b_f=r_f$ and the same
continuous semi-norm $p$ as in (\ref{ZO1Z}).

\medskip

We denote by $\Omega^{{\bf C}}$ the complexification of $\Omega:
\Omega^{{\bf C}}= \Omega\oplus i \Omega.$ We denote by ${\cal
V}(\Omega)$ the class of functions $f: \Omega\to {\bf R}, f(0)=0,$
which can be analytically continued onto $\Omega^{{\bf C}}$ and they
have the exponential growth.

\medskip

{\bf Lemma 6.2.} {\it Let $\rho\in S_G^\alpha (\Omega).$ Then, for
any function $f \in {\cal V}(\Omega),$ the following asymptotic
equality holds:
\begin{equation}
\label{BER}
 <f>_\rho \equiv  \int_\Omega f(\psi) d \rho
(\psi)=\frac{\alpha}{2} \int_\Omega (f^{\prime\prime} (0) \psi,
\psi) d\rho_D (u) + o(\alpha), \; \alpha \to 0,
\end{equation} where $D=\frac{{\rm cov}
\rho}{\alpha}.$ Here
\begin{equation}
\label{OL} o(\alpha) = \alpha^2 R(\alpha, f, \rho),
\end{equation}
where
\begin{equation}
\label{OL1} \vert R(\alpha,f,\rho)\vert \leq c_f\int_\Omega  e^{r_f
p(\psi)}d\rho_D (\psi).
\end{equation}
The semi-norm $p$ is determined by the inequality (\ref{WS}).}

The proof of this Theorem repeats the proof of Lemma 3.2. Instead of
Lemma 3.1, we apply its generalization to the case of an arbitrary
locally convex topological linear space, see Lemma 6.1.

\medskip

 We pay attention that $D: \Omega^\prime \to \Omega,$ and
$A=\frac{f^{\prime\prime (0)}}{2}: \Omega \to \Omega^\prime,$ so
$C=DA: \Omega \to \Omega.$ In general, this operator can not be
extended to a continuous operator in $H.$ We would like to obtain an
analogue of the formula (\ref{QI}) for linear continuous operators
$A: \Omega \to \Omega^\prime:$
\begin{equation}
\label{EF} \int_\Omega (A \psi, \psi) d \rho_D (u)={\rm Tr}\; DA
\end{equation}
The main mathematical problem is that in general the operator $C=DA$
is not even continuous in $H$, so it is not a trace class operator
in the Hilbert space $H.$ Nevertheless, we can introduce the notion
of trace even in such a framework.

We recall that systems of vectors $\{e_j\}_{j=1}^\infty, e_j \in
\Omega,$ and $\{e_j^\prime\}^\infty_{j=1}, e_j^\prime \in
\Omega^\prime,$ are called {\it biorthogonal topological bases} in
$\Omega$ and $\Omega^\prime$ if
$$
(e_j^\prime, e_i)=\delta_{ij},\; \mbox{and}\; \psi=\sum_{j=1}^\infty
(e_j^\prime, \psi) e_j, \psi \in \Omega, \phi=\sum_{j=1}^\infty
(\phi, e_j) e_j^\prime, \phi \in \Omega^\prime,
$$
where the series converge in $\Omega$ and $\Omega^\prime, $
respectively.

{\bf Definition 6.1.} {\it A linear continuous operator $C: \Omega
\to \Omega$ is called trace-class operator if, for any pair of
biorthogonal topological bases, the series
$${\rm Tr} \; C=\sum_{j=1}^\infty
(e_j^\prime, C e_j)$$ converges and its sum does not depend on
bases.}

{\bf Lemma 6.3.} {\it Let $\rho$ be a Gaussian measure on $\Omega$
and let $A: \Omega \to \Omega^\prime$ be a continuous operator. Then
the operator $C=DA,$ where $D ={\rm cov} \rho,$ belongs to the trace
class and the equality (\ref{EF}) holds.}

As a consequence of Lemmas 6.2 and 6.3, we obtain:

{\bf Theorem 6.1.} {\it Let $\rho\in S_G^\alpha (\Omega).$ Then, for
any function $f \in {\cal V}(\Omega),$ the following asymptotic
equality holds:
\begin{equation}
\label{BERZ}
 <f>_\rho \equiv  \int_\Omega f(\psi) d \rho
(\psi)= \rm{Tr} \; D f^{\prime\prime}(0)/2  + o(\alpha), \; \alpha
\to 0,
\end{equation} where $D=\frac{{\rm cov}
\rho}{\alpha}.$}

\medskip

 Thus our prequantum model, PCSFT, provides the motivation to extend the set of
 quantum observables and consider all continuous operators $A: \Omega \to \Omega^\prime.$
 Operators should be  self-adjoint in the ordinary sense:
 $A^\prime=A.$  We recall that here
 $A^\prime: \Omega^{\prime\prime}\to \Omega^\prime,$ but $\Omega^{\prime\prime} \equiv \Omega,$
 since $\Omega$ is a nuclear Frechet space and hence it is reflexive.  Denote the set of such
 operators by the symbol ${\cal L}_s(\Omega, \Omega^\prime).$ Denote the set
 of covariation operators of Gaussian measures belonging the space
 $S_G^1(\Omega)$ by the symbol ${\cal D}(\Omega^\prime, \Omega).$

 {\bf Definition 6.2.} {\it A statistical quantum model corresponding
 to a rigged Hilbert space ${\cal T}$ given by (\ref{GT}) is the pair
 $$
 N_{\rm{quant}}({\cal T})= ({\cal D}(\Omega^\prime, \Omega),{\cal L}_s(\Omega,
 \Omega^\prime)).
 $$
 A generalized density operators $D\in {\cal D}(\Omega^\prime, \Omega)$
 represents a statistical state; a linear operator $A\in {\cal L}_s(\Omega,
 \Omega^\prime)$ represents a quantum observable. The average of such an
 observable with respect to such a statistical state is given by the
 following generalization of the von Neumann trace-formula:}
 \begin{equation}
 \label{RT}
 <A>_D= \rm{Tr} \; D A
\end{equation}

We choose the  state space $\Omega$ -- a nuclear Frechet space. For
a rigged Hilbert space ${\cal T}$ given by (\ref{GT}), we consider
the classical statistical model $M^\alpha({\cal
T})=(S_G^\alpha(\Omega), {\cal V}(\Omega)).$ Here as always
$<f>_\rho= \int_\Omega f(\psi) d\rho(\psi).$

The equality (\ref{BERZ}) can be used as the motivation for defining
the following classical $\to$ quantum map $T$ from the classical
statistical model $M^\alpha({\cal T})=(S_G^\alpha(\Omega), {\cal
V}(\Omega))$ onto the quantum statistical model
$N_{\rm{quant}}({\cal T})= ({\cal D}(\Omega^\prime, \Omega), {\cal
L}_s(\Omega, \Omega^\prime))$ by (\ref{Q20}), (\ref{Q30}). Our
previous considerations can be presented as

\medskip

{\bf Theorem 6.2.} {\it The map $T: S_G^\alpha(\Omega) \to {\cal
D}(\Omega^\prime, \Omega)$ is one-to-one; the map $T: {\cal
V}(\Omega) \to {\cal L}_s(\Omega, \Omega^\prime)$ is linear
surjection and the classical and quantum averages are coupled by the
asymptotic equality (\ref{BERZ}).}

\medskip

{\bf Example 6.2.} The position operators $\hat x_j, j=1, 2, 3$ can
be obtained as $\hat x_j=\frac{1}{2} f_{x_j}^{\prime\prime}(0),$
where
$$
f_{x_j}(\psi)= \int_{{\bf R}^3} x_j \psi^2(x)  dx.
$$
Here the operator of multiplication $\hat x_j: {\cal S} ({\bf R}^3)
\to {\cal S} ({\bf R}^3), \psi \to x_j \psi,$ is continuous. Hence
$\hat x_j: {\cal S} ({\bf R}^3) \to {\cal S}^\prime ({\bf R}^3)$ is
also continuous. Thus, for any measure $\rho \in S_G^\alpha ({\cal
S} ({\bf R}^3)),$ we have
$$
<f_{x_j}>_\rho \equiv  \int_{{\cal S}({\bf R}^3)} \int_{{\bf R}^3} x
\psi^2 (x) dx d \rho(\psi) = \alpha {\rm Tr} \; D \hat x_j,
$$
$D=\rm{cov} \; \rho/\alpha$  (here the trace of the composition $D
\hat x_j$ is well defined).

{\bf Example 6.3.} Let $x_0$ be a fixed point in ${\bf R}^3.$ Let
now $A \psi(x)=\delta (x-x_0) \psi (x),  \psi \in {\cal S} ({\bf
R}^3).$ This operator does not belong to the domain of the
conventional quantum formalism. It could not be represented as an
unbounded operator in $H=L_2 ({\bf R}^3)$ with a dense domain of
definition.
 Nevertheless,
 $$\int_{{\cal S} ({\bf R}^3)}(A \psi,  \psi) d\rho (\psi)
 =\int_{{\cal S} ({\bf R}^3)} \psi^2 (x_0) d\rho ( \psi)=
 \alpha {\rm Tr} \; D A
 $$
 and the trace of the composition $D A$ is well defined.

{\bf Example 6.4.} The momentum operators $\hat p_j, j=1,2,3,$ can
be obtained as $\hat p_j=\frac{1}{2} f_{p_j}^{\prime\prime}(0),$
where
$$f_{p_j}(\psi)= - i \int_{{\bf R}^3} \frac{\partial \psi}{\partial x_j}(x)
\overline{\psi(x)} dx.
$$
Here the operator $\hat p_j: {\cal S} ({\bf R}^3) \to {\cal S}({\bf
R}^3)$ is continuous. Hence, $\hat p_j: {\cal S}({\bf R}^3) \to
{\cal S}^\prime ({\bf R}^3)$ is also continuous. Thus for any
measure $\rho \in S_{G,{\rm symp}}^\alpha({\cal S} ({\bf R}^3)
\times {\cal S}({\bf R}^3)), $ we have (for $\psi(x)=q(x) + ip
(x)$):
$$
<f_{p_j}>_\rho\equiv  -i \int_{{\cal S}({\bf R}^3) \times {\cal S}
({\bf R}^3)} \int \frac{\partial \psi}{\partial x_j}(x)
\overline{\psi(x)} dx d \rho (\psi)=\alpha {\rm Tr} \; D \; \hat
p_j,
$$
where $D=\rm{cov} \; \rho/\alpha.$ Here $D\hat p_j: {\cal S}({\bf
R}^3) \to {\cal S}({\bf R}^3)$ is the trace class operator. Similar
considerations can be done for angular momentum operators.

\section{Generalized quantum mechanics: approximations of higher orders}

We have created the classical statistical model which induced the
quantum statistical model. The quantum  description can be obtained
through the Taylor expansion of classical physical variables up to
the terms of the second order. The crucial point is the presence of
a parameter $\alpha$ which small in QM, but not in the prequantum
classical model.

This viewpoint to conventional  quantum mechanics implies the
evident possibility to generalize this formalism by considering
higher orders of the  Taylor expansion of classical physical
variables and corresponding expansions of classical averages with
respect to the parameter $\alpha.$

We still consider the real case: $\Omega=H,$ where $H$ is the real
separable Hilbert space, and only bounded linear operators (and
forms).
 We recall that momentums of a measure $\rho$ are defined by
$$
a_\rho^{(k)} (z_1, \ldots, z_k)=\int_\Omega (z_1,  \psi)...(z_k,
\psi) d \rho ( \psi).
$$
In particular, $a_\rho^{(1)}\equiv a_\rho$ is the mean value and
$a_\rho^{(2)}$ is the covariation form.  We remark that for a
Gaussian measure $\rho, a_\rho=0$ implies that all its momenta of
odd orders $a_\rho^{(k)}, k=2n + 1, n=0, 1, \ldots, $ are also equal
to zero.

Therefore the expansion of $<f>_\rho$ with respect to
$s=\alpha^{1/2}$ does not contain terms with $s^{2n + 1}.$ Hence
this is the expansion with respect to $\alpha^n(=s^{2n}), n=1,2,
\ldots$ We are able to create $o(\alpha^n)$-generalization of
quantum mechanics through neglecting by terms of the magnitude
$o(\alpha^n), \alpha \to 0 (n=1,2, \ldots)$ in the power expansion
of the classical average. Of course, for $n=1$ we obtain the
conventional quantum mechanics.  Let us consider the classical
statistical model
 \begin{equation}
\label{TRHJ6} M^\alpha=(S_G^\alpha (\Omega), {\cal V}(\Omega)).
\end{equation}
 By taking into account
that $a_\rho^{2n + 1}=0, n=0, 1, \ldots, $ for $\rho \in
S_G^\alpha(\Omega),$ we have:
\begin{equation}
\label{GI2} <f>_\rho=\frac{\alpha}{2} {\rm Tr} \; D f^{\prime
\prime} (0) + \sum_{k=2}^\infty \frac{\alpha^k}{(2k)!} \int_\Omega
f^{(2k)} (0) (\phi, \ldots, \phi) d \rho_D (\phi),
\end{equation}
where as always $D=\frac{{\rm cov} \rho}{\alpha}.$

We now consider a new epistemic (``observational'') statistical
model which is a natural generalization of the conventional quantum
mechanics. We start with some preliminary mathematical
considerations. Let $A$ and $B$ be two $n$-linear symmetric forms.
We define their trace by
\begin{equation}
\label{QPJ6} {\rm Tr} \;  B A=\sum_{j_1, \ldots, j_n=1}^\infty
B(e_{j_1}, \ldots, e_{j_h}) A(e_{j_1}, \ldots, e_{j_n}),
\end{equation}
if this series converges and its sum does not depend on the choice
of an orthonormal basis $\{e_j\}$ in $\Omega.$ We remark that
\begin{equation}
\label{H} <f>_\rho=\frac{\alpha}{2} {\rm Tr} \; D f^{\prime \prime}
(0) + \sum_{k=2}^n \frac{\alpha^k}{2k!} {\rm Tr} \;a_{\rho_D}^{(2k)}
f^{(2k)} (0) + o(\alpha^n), \alpha \to 0,
\end{equation}
Here we used the following  result about Gaussian integrals:

{\bf Lemma 7.1.} {\it Let $A_k$ be a continuous $k$-linear form on
$\Omega$ and let $ \rho_D$ be a Gaussian measure (with zero mean
value and the covariation operator $D).$ Then}
\begin{equation}
\label{AKH} \int_\Omega A_k (\psi, \ldots, \psi) d \rho_D
(\psi)={\rm Tr} \; a_{\rho_D}^{(k)} A_k.
\end{equation}

{\bf Proof.} Let $\{e_j\}_{j=1}^\infty$ be an orthonormal basis in
$\Omega.$ We apply the well known Lebesque theorem on majorant
convergence. We set
\begin{equation}
\label{QPJ5} f_N(\psi)=\sum_{j_1, \ldots, j_k=1}^n A_k(e_{j_1},
\ldots, e_{j_k}) (e_{j_1}, \psi) \ldots (e_{j_k}, \psi).
\end{equation}
We have
\begin{equation}
\label{QPJ4}|f_N(\psi)|=|A_k (\sum_{j_1=1}^N (x, e_{j_1}) e_{j_1}
\ldots, \sum_{j_k=1}^N (\psi, e_{j_k}) e_{j_k})|\leq||A_k|| \;
||\psi||^k.
\end{equation}
Therefore we obtain:
$$
\int_\Omega A_k (\psi, \ldots, \psi) d\rho_D(\psi)= \lim_{N\to
\infty} \int_\Omega f_N(\psi) d\rho_D(\psi)
$$
\begin{equation}
\label{QPJ13} = \sum_{j_1=1,..., j_k=1}^\infty A_k (e_{j_1}, \ldots,
e_{j_k}) \int_\Omega (e_{j_1}, \psi)\ldots (e_{j_k}, \psi)
d\rho_D(\psi)={\rm Tr}\; a^{(k)}_{\rho_D} A_k.
\end{equation}
The proof is finished.
\bigskip

In particular, we obtained the following inequality:
\begin{equation}
\label{QPJ2} \vert {\rm Tr} \; a^k_{\rho_D} A_k|\leq
||A||\int_\Omega||\psi||^k d\rho_D (\psi).
\end{equation}

We now remark that for a Gaussian measure (with zero mean value)
integrals (\ref{AKH}) are equal to zero for  $k=2l+1.$   Thus ${\rm
Tr} \; a^{(2l+1)}_{\rho_D} A_{2l+1}=0.$  It is easy to see that
$2k$-linear forms (momenta of even order) $a_{\rho_D}^{2k}$ can be
expressed through the covariance operator $D:$
\begin{equation}
\label{QPJ1} a_{\rho_D}^{(2k)}=e(k,
D)=\frac{d^{2k}}{d\phi^{2k}}e^{-\frac{1}{2}(D\phi, \phi)}|_{\phi=0}.
\end{equation}
In particular, $e(2, D)(\phi_1, \phi_2)=(D\phi_1, \phi_2)$ and $e(4,
D)(\phi_1, \phi_2, \phi_3, \phi_4)$
\begin{equation}
\label{QPJ} = (D\phi_1, \phi_3)(D\phi_2, \phi_4) + (D\phi_2,
\phi_3)(D\phi_1, \phi_4) + (D\phi_1, \phi_2)(D\phi_3, \phi_4).
\end{equation}
Thus (\ref{H}) can be rewritten as
\begin{equation}
\label{EH} <f>_{\rho_B}= \frac{\alpha}{2} {\rm Tr} \;
Df^{\prime\prime}(0) + \sum_{k=2}^n\frac{\alpha^k}{2k!} {\rm Tr} \;
e(2k, D)f^{(2k)}(0) + o(\alpha^n), \; \alpha\to 0,
\end{equation}
or by introducing the $1/\alpha$-amplification of the classical
physical variable $f$ we have:
\begin{equation}
\label{EHA} <f_\alpha>_{\rho_B}=\frac{1}{2} {\rm Tr} \;
Df^{\prime\prime}(0) + \sum_{k=2}^n\frac{\alpha^{k-1}}{2k!} {\rm Tr}
\; e(2k, D)f^{(2k)}(0) + o(\alpha^{n-1})
\end{equation}

This formula is the basis of {\it a new quantum theory.} In this
theory statistical states can be still represented by von Neumann
density operators $D \in {\cal D} (\Omega),$ but observables are
represented by
 multiples $A=(A_2, A_4, \ldots, A_{2n}),$ where $A_{2j}$
 are  symmetric $2n$-linear forms on a Hilbert space $\Omega.$ In particular, the
 quadratic form $A_2$ can be represented by a self-adjoint operator. To escape mathematical
 difficulties, we can assume that forms $A_{2j}$ are continuous. Denote the space
 of all such multiples $A$ by $L_{2n}(\Omega).$ We obtain the following generalization of the conventional quantum model:
\begin{equation}
\label{EHJ7} N_{\rm{quant},2n}= ({\cal D}(\Omega), L_{2n}(\Omega)).
\end{equation}
 Here the average of
an observable $A\in L_{2n}(\Omega)$ with respect to a state $D\in
{\cal D}(\Omega)$ is given by
\begin{equation}
\label{F} <a>_D=\sum_{n=1}^n {\rm Tr} \; e(2k, D) A_{2k}
\end{equation}
If one define ${\rm Tr} \; D A=\sum_{k=1}^n {\rm Tr} \; e(2k, D)
A_{2k}$ then the formula (\ref{F}) can be written as in the
conventional quantum mechanics (von Neumann's formula of $n$th
order):
\begin{equation}
\label{V} <A>_D={\rm Tr}D A
\end{equation}
This model is the result of the following ``quantization'' procedure
of the classical statistical model $M^\alpha=(S_G^\alpha (\Omega),
{\cal V}(\Omega)$:
\begin{equation}
\label{QP} \rho \to D=\frac{{\rm cov} \rho}{\alpha};
\end{equation}
\begin{equation}
\label{QP1} f\to A=(\frac{1}{2} f^{\prime\prime}(0),
\frac{\alpha}{4!}f^{(4)}(0),..., \frac{\alpha^{n-1}}{(2n)!}
f^{(2n)}(0)).
\end{equation}
(thus here $A_{2k}=\frac{\alpha^{k-1}}{(2k)!} f^{(2k)}(0)).$ The
transformation $T_{2n}$ given by (\ref{QP}), (\ref{QP1}) maps the
classical statistical model $M^\alpha=(S_G^\alpha (\Omega), {\cal
V}(\Omega))$ onto generalized quantum model $N_{\rm{quant},2n}=(
{\cal D} (\Omega), L_{2n} (\Omega)).$

\medskip

{\bf Theorem 7.1.} {\it For the classical statistical model
$M^\alpha=(S_G^\alpha(\Omega),$\\ ${\cal V}(\Omega))$, the classical
$\to$ quantum map $T_{2n},$ defined by  (\ref{QP}) and (\ref{QP1}),
is one-to-one for statistical states; it has a huge degeneration for
variables. Classical and quantum averages are coupled through the
asymptotic equality (\ref{EH})}.

\medskip

We pay attention to the simple mathematical fact that the degree of
degeneration of the map $T_{2n}: {\cal V}(\Omega)\to L_{2n}(\Omega)$
is decreasing for $n \to \infty.$ Denote the space of polynomials of
the degree $2n$ containing only terms of even degrees by the symbol
$P_{2n}.$ Thus $f \in P_{2n}$ iff $f(\psi)=Q_2(\psi, \psi) + Q_4
(\psi, \psi, \psi, \psi) + \ldots + Q_{2n} (\psi, \ldots ,\psi),$
where $Q_{2j}:\Omega^{2j} \to {\bf R}$ is a symmetric $2j$-linear
(continuous) form. The restriction of the map $T_{2n}$ on the
subspace $P_{2n}$ of the space ${\cal V}$ is one-to-one. One can
also consider a generalized quantum model
\begin{equation}
\label{QPY} N_{\rm{quant},\infty}=({\cal D}, L_\infty),
\end{equation}
where $L_\infty(\Omega)$ consists of infinite sequences of
$2n$-linear (continuous) forms on $\Omega:$
\begin{equation}
\label{QPY1} A=(A_2, \ldots, A_{2n}, \ldots).
\end{equation}
The correspondence between the classical model $M^\alpha$ (for any
$\alpha$) and the generalized quantum model $ N_{\rm{quant},\infty}$
is one-to-one.

This paper was partially supported by EU-network "Quantum
Probability and Applications."

\bigskip

{\bf References}

[1]  D. Hilbert, J. von Neumann, L. Nordheim, {\it Math. Ann.}, {\bf
98}, 1-30 (1927).

[2] P. A. M.  Dirac, {\it The Principles of Quantum Mechanics,}
Oxford Univ. Press, 1930.

[3]  W. Heisenberg, {\it Physical principles of quantum theory,}
Chicago Univ. Press, 1930.

[4] J. von Neumann, {\it Mathematical foundations of quantum
mechanics,} Princeton Univ. Press, Princeton, N.J., 1955.

[5] A. Einstein, B. Podolsky, N. Rosen, {\it Phys. Rev.} {\bf 47},
777--780 (1935).

[6] E. Schr\"odinger,  {\it Philosophy and the Birth of Quantum
Mechanics.} Edited by M. Bitbol, O. Darrigol (Editions Frontieres,
Gif-sur-Yvette, 1992); especially the paper of S. D'Agostino,
``Continuity and completeness in physical theory: Schr\"odinger's
return to the wave interpretation of quantum mechanics in the
1950's'', pp. 339-360.

[7]   E. Schr\"odinger, {\it E. Schr\"odinger Gesammelte
Abhandlungen} ( Wieweg and Son, Wien, 1984); especially the paper
``What is an elementary particle?'', pp. 456-463.

[8] A. Einstein, {\it The collected papers of Albert Einstein}
(Princeton Univ. Press, Princeton, 1993).

[9] A. Einstein and L. Infeld, {\it The evolution of Physics. From
early concepts to relativity and quanta} (Free Press, London, 1967).

[10]  A. Lande, {\it New foundations of quantum mechanics,}
Cambridge Univ. Press, Cambridge, 1965.

[11] L. De Broglie, {\it The current interpretation of wave
mechanics, critical study.} Elsevier Publ., Amsterdam-London-New
York, 1964.

[12] J. S. Bell, {\it Speakable and unspeakable in quantum
mechanics,} Cambridge Univ. Press, 1987.

[13] G. W. Mackey, {\it Mathematical foundations of quantum
mechanics,} W. A. Benjamin INc, New York, 1963.

[14]  S. Kochen and E. Specker, {\it J. Math. Mech.}, {\bf 17},
59-87 (1967).

[15] L. E. Ballentine, {\it Rev. Mod. Phys.}, {\bf 42}, 358--381
(1970).

[16]  G. Ludwig, {\it Foundations of quantum mechanics,} Springer,
Berlin, 1983.

[17] E. B. Davies, J. T. Lewis, {\it Comm. Math. Phys.} {\bf 17},
239-260 (1970).

[18] E. Nelson, {\it Quantum fluctuation,} Princeton Univ. Press,
Princeton, 1985.

G.C. Ghirardi, C. Omero, A. Rimini and T. Weber, The Stochastic
Interpretation of Quantum Mechanics: a Critical Review, {\it Rivista
del Nuovo Cimento} {\bf 1} 1 (1978).

S. Albeverio, and R. H\"oegh-Krohn, A remark on the connection
between stochastic mechanics and the heat equation. {\it J. Math.
Phys.}, {\bf 15}, 1745-1747 (1975).

J. L\"orinczi, R. A. Minlos, and  H. Spohn, The infrared behaviour
in Nelson's model of a quantum particle coupled to a massive scalar
field. {\it. Ann. H. Poincare}, {\bf 3}, 269-295 (2002).

[19] D.  Bohm  and B. Hiley, {\it The undivided universe: an
ontological interpretation of quantum mechanics,} Routledge and
Kegan Paul, London, 1993.

[20] S. P. Gudder, {\it Trans. AMS} {\bf 119}, 428-442 (1965).

[21] S. P. Gudder, {\it Axiomatic quantum mechanics and generalized
probability theory,} Academic Press, New York, 1970.

[22] H. Spohn, Quantum measurement theory including initial
correlations and observables with continuous spectrum. {\it Int. J.
Theor. Phys.}, {\bf 15}, 283-375 (1976).

[23] R. Feynman and A. Hibbs, {\it Quantum Mechanics and Path
Integrals,} McGraw-Hill, New-York, 1965.

[24] J. M. Jauch, {\it Foundations of Quantum Mechanics,}
Addison-Wesley, Reading, Mass., 1968.

[25] A. Peres, {\em Quantum Theory: Concepts and Methods,}
Dordrecht, Kluwer Academic, 1994.

[26] L. Accardi, {\it ``The probabilistic roots of the quantum
mechanical paradoxes''} in {\em The wave--particle dualism.  A
tribute to Louis de Broglie on his 90th Birthday,} edited by  S.
Diner, D. Fargue, G. Lochak and F. Selleri, D. Reidel Publ. Company,
Dordrecht, 1984, pp. 297--330.

[27] L. Accardi, {\it Urne e Camaleoni: Dialogo sulla realta, le
leggi del caso e la teoria quantistica,} Il Saggiatore, Rome, 1997.

[28]  L. E. Ballentine, {\it Quantum mechanics,} Englewood Cliffs,
New Jersey, 1989.

[29] L. E. Ballentine,  {\it ``Interpretations of probability and
quantum theory'',} in {\it Foundations of Probability and Physics,}
edited by  A. Yu. Khrennikov,
 Q. Prob. White Noise Anal.,  13,  WSP, Singapore, 2001, pp. 71-84.

[30] A. S. Holevo, {\it Probabilistic and statistical aspects of
quantum theory,} North-Holland, Amsterdam,  1982.

[31] A. S. Holevo, {\it Statistical structure of quantum theory,}
Springer, Berlin-Heidelberg, 2001.

[32] P. Busch, M. Grabowski, P. Lahti, {\it Operational Quantum
Physics,} Springer Verlag,Berlin, 1995.

[33] A. Yu. Khrennikov (editor), {\it Foundations of Probability and
Physics,} Q. Prob. White Noise Anal.,  13,  WSP, Singapore, 2001.

[34] A. Yu. Khrennikov (editor), {\it Quantum Theory:
Reconsideration of Foundations,} Ser. Math. Modeling, 2, V\"axj\"o
Univ. Press,  2002.

[35] A. Yu. Khrennikov (editor), {\it Foundations of Probability and
Physics}-2, Ser. Math. Modeling, 5, V\"axj\"o Univ. Press,  2003.

[36] A. Yu. Khrennikov (editor),  {\it Quantum Theory:
Reconsideration of Foundations}-2,  Ser. Math. Modeling, 10,
V\"axj\"o Univ. Press,  2004.

[37]  A. Yu. Khrennikov (editor),  Proceedings of Conference {\it
Foundations of Probability and Physics-3,} American Institute of
Physics, Ser. Conference Proceedings, {\bf 750}, 2005.

[38] A. Yu. Khrennikov, {\it Interpretations of Probability,} VSP
Int. Sc. Publishers, Utrecht/Tokyo, 1999 (second edition, 2004).

[39] A. E. Allahverdyan, R. Balian, T. M. Nieuwenhuizen, in: A. Yu.
Khrennikov (Ed.), Foundations of Probability and Physics-3,
Melville, New York: AIP Conference Proceedings, 2005, pp. 16-24.

[40] W. De Baere,  {\it Lett. Nuovo Cimento} {\bf 39}, 234 (1984);
{\bf 40}, 488(1984); {\it Advances in electronics and electron
physics} {\bf 68}, 245 (1986).

[41]  De Muynck W. M., {\it Foundations of Quantum Mechanics, an
Empiricists Approach} (Kluwer, Dordrecht) 2002.

[42] De Muynck W., De Baere W., Martens H., {\it Found. of Physics}
{\bf 24} (1994) 1589.

[43] Allahverdyan A. E., Balian R., Nieuwenhuizen Th., {\it
Europhys. Lett.} {\bf 61} (2003) 452.

[44] K. Hess and W. Philipp, {\it Proc. Nat. Acad. Sc.} {\bf 98},
14224 (2001); {\bf 98}, 14227(2001); {\bf 101}, 1799 (2004); {\it
Europhys. Lett.} {\bf 57}, 775 (2002).

[45] A. Yu. Khrennikov, {\it J. Phys.A: Math. Gen.} {\bf 34},
9965-9981 (2001); {\it Il Nuovo Cimento} {\bf B 117},  267-281
(2002); {\it J. Math. Phys.} {\bf 43}, 789-802 (2002); {\it
Information dynamics in cognitive, psychological and anomalous
phenomena,} Ser. Fundamental Theories of Physics, Kluwer, Dordreht,
2004;  {\it J. Math. Phys.} {\bf 44},  2471- 2478 (2003); {\it Phys.
Lett. A} {\bf 316}, 279-296 (2003); {\it Annalen  der Physik} {\bf
12},  575-585 (2003).

[46] A. Yu. Khrennikov, A pre-quantum classical statistical model
with infinite-dimensional phase space. {\it J. Phys. A: Math. Gen.},
{\bf 38}, 9051-9073 (2005).

[47] A. Yu. Khrennikov, Quantum mechanics as an asymptotic
projection of statistical mechanics of classical fields: derivation
of Schr\"odinger's, Heisenberg's and von Neumann's equations.
http://www.arxiv.org/abs/quant-ph/0511074

[48]  O. G. Smolyanov,  Infinite-dimensional pseudodifferential
operators and Schr\"odinger quantization. {\it Dokl. Akad. Nauk
USSR}, {\bf 263}, 558(1982).

[49] A. Yu. Khrennikov, Infinite-dimensional pseudo-differential
operators. {\it Izvestia Akademii Nauk USSR, ser.Math.}, {\bf 51},
46 (1987).

[50] L. de la Pena and A. M. Cetto, {\it The Quantum Dice: An
Introduction to Stochastic Electrodynamics Kluwer.} Dordrecht, 1996;
T. H. Boyer, {\it A Brief Survey of Stochastic Electrodynamics} in
Foundations of Radiation Theory and Quantum Electrodynamics, edited
by A. O. Barut, Plenum, New York, 1980; T. H. Boyer, Timothy H.,
{\it Scientific American},pp 70-78, Aug 1985; see also an extended
discussion on vacuum fluctuations in: M. O. Scully, M. S. Zubairy,
{\it Quantum optics,} Cambridge University Press, Cambridge, 1997;
W. H. Louisell, {\it Quantum Statistical Properties of Radiation.}
J. Wiley, New York, 1973; L. Mandel and E. Wolf, {\it Optical
Coherence and Quantum Optics.} Cambridge University Press,
Cambridge, 1995.

[51] L. De La Pena, {\it Found. Phys.} {\bf 12}, 1017 (1982); {\it
J. Math. Phys.} {\bf 10}, 1620 (1969); L. De La Pena, A. M. Cetto,
{\it Phys. Rev. D} {\bf 3}, 795 (1971).

[52] A. Bach,  {\it J. Math. Phys.} {\bf 14} 125 (1981).

[53] A. Bach, {\it Phys. Lett. A} {\bf 73} 287 (1979).

[54] A. Bach,  {\it J. Math. Phys.} {\bf 21} 789 (1980).

[55] A. V. Skorohod, Integration in Hilbert space. Springer-Verlag,
Berlin, 1974.

[56] T. Hida,  {\it Selected Papers of Takeyiki Hida.} H. H. Kuo, N.
Obata, K. Saito, L. Streit, Si Si, L. Accardi (editors) World
Scientific Publ. (2001).

[57] T. Hida, M. Hitsuda {\it Gaussian Processes,} Translations of
Mathematical Monographs, {\bf 120}, American Mathematical Society,
1993.

[58] S. Albeverio, and M. R\"ockner, {\it Prob. Theory and Related
Fields} {\bf 89}, 347 (1991).

S. Albeverio, R. H\"oegh-Krohn, Dirichlet forms and diffusion
processes on rigged Hilbert spaces. {\it  Zeitschrift f\"ur
Wahrscheinlichketstheorie und verwandte Gebite}, {\bf 40}, 59-106
(1977).

[59] B. Simon, {\it Functional Integration and Quantum Physics.} Ams
Chelsea Pub., 2005)

[60] A. Yu. Khrennikov, Equations with infinite-dimensional
pseudo-differential operators. Dissertation for the degree of
candidate of phys-math. sc., Dept. Mechanics-Mathematics, Moscow
State University, Moscow, 1983.

[61] O. G. Smolyanov and S. V. Fomin,  Measures on topological
linear spaces. {\it Russian Math. Surveys}, {\bf 31}, 3-5 (1976).

\end{document}